\documentclass[%
 twocolumn,
 amsmath,amssymb,
 aps,
pre,
]{revtex4-2}

\usepackage{graphicx}
\usepackage{dcolumn}
\usepackage{bm}

\def\<{\langle}
\def\>{\rangle}

\def\g2{g^{(2)}}

\newcommand{\non}{\nonumber}
\newcommand{\be}{\begin{equation}}
\newcommand{\ee}{\end{equation}}
\newcommand{\bea}{\begin{eqnarray}}
\newcommand{\eea}{\end{eqnarray}}

\begin{document}

\preprint{}

\title{On the definition of heat current for periodic systems and its implications for simulations of thermal conductivity in solids}

\author{Andrey Pereverzev}
 \email{pereverzeva@missouri.edu}
\affiliation{%
 Department of Chemistry, University of Missouri,\\ Columbia, Missouri 65211-7600, USA
}%


\date{\today}

\begin{abstract}
We re-derive the expression for the heat current for a classical system subject to periodic boundary conditions
and show that it can be written as a sum of two terms. The first term is a time derivative of the first moment of the system energy density while the second term is expressed 
through the energy transfer rate through the periodic boundary. 
We show that in solids
the second term alone leads to the same thermal conductivity 
as the full expression for the heat current when used in the Green-Kubo approach. More generally, energy passing 
though  any surface formed by translation of the original periodic boundary can be used to calculate thermal conductivity.
These statements are verified for two systems: crystalline argon and crystals of argon and krypton forming an interface.
\end{abstract}

\maketitle

\section{Introduction}
The correct definition of heat current is crucial both for fundamental understanding of heat transport mechanism in various materials and for numerical calculations of thermal conductivity.
In particular, calculation of the thermal conductivity tensor $\kappa_{\alpha\beta}$ using classical molecular dynamics can be accomplished using the Green-Kubo  formalism \cite{Kubo,Mahan} in which 
\begin{equation}
\kappa_{\alpha\beta}=\frac{1}{k_BT^2V}\int_0^\infty dtC_{\alpha\beta}(t), \label{qcurr}
\end{equation}
where  $T$ is temperature, $V$ is the system volume, and
\be
C_{\alpha\beta}(t)=\langle J_{\alpha}(0) J_{\beta}(t)\rangle \label{corrfun}
\ee
is the tensoral time correlation function in which
 $J_{\alpha}$ represents the $\alpha$th component of the heat current
 and the brackets indicate averaging over an equilibrium ensemble.

The Green-Kubo approach has become a powerful tool for obtaining thermal conductivities of various materials due, in part, to the simplicity of its implementation, which only involves equilibrium molecular dynamics simulations \cite{Ladd,Che,McGaughey1,McGaughey2,Izvekov,Landry,McGaughey3,Henry,Yip,Fan}.
The main numerical disadvantage of the Green-Kubo approach as implemented in molecular dynamics is a slow convergence of the integral in Eq. (\ref{qcurr}) which  may require very long simulation times and careful analysis of the correlation function \cite{Yip,Li,McGaughey2,PereverzevSewell}.

 The general definition of the heat current was 
first given by Hardy \cite{Hardy} for nonperiodic systems. However, the majority of molecular dynamics simulations are performed using periodic
boundary conditions and  Hardy's definition has to be properly adjusted for
Eq. (\ref{qcurr}) to give the correct results.
The expression for the heat current that accounts for the  periodic boundaries has been in use in molecular dynamics for quite some time {\cite{Ladd,Che,McGaughey1,McGaughey2,Izvekov,Landry,McGaughey3,Henry,Yip,Fan,Surblys,Plimpton}}. 
In spite of its wide use, a clear derivation of such expression for systems with arbitrary multi-body potentials is hard to find.
For example, Ref. \cite{Todd} provides a detailed derivation of the pressure tensor for periodic systems with pair-wise interactions and 
states that the heat current for such systems can be obtained similarly. By contrast, Refs.\cite{Fan,Surblys,Boone} give derivations of the heat current for the
many-body potential but do not discuss explicitly how the periodic boundaries are taken into account.

In this work we re-derive the expression for heat current of a periodic system in which the boundary effect is clearly identified and then, using 
its functional form,  show that 
it leads to some surprising theoretical 
and numerical implications for the heat transport in solids.
These implications are based on the recently discovered invariance of 
$\kappa_{\alpha\beta}$ given by Eq. (\ref{qcurr}) to addition of 
the time derivative of a function to the heat current \cite{Baroni,Baroni2,Ercole,Ercole2016}.

\section{Heat current expression for a periodic system} \label{secone}

A conventional way to treat periodic boundary conditions is to assume that the system of interest which is placed in the central parallelepiped (or triclinic) box 
 is surrounded by an infinite number of translated identical image boxes that fill up the space \cite{Erpenbeck,Todd,Bekker,Thompson}.
It is convenient to specify these boxes with 
vector ${\bf n}$ given by
\be
{\bf n}=\eta_a{\bf a}+\eta_b{\bf b}+\eta_c{\bf c}, \label{vector}
\ee
where ${\bf a}$, ${\bf b}$, and ${\bf c}$ are the vectors specifying the three edges of the central box and 
 $\eta_a$, $\eta_b$, and $\eta_c$ can take any integer values. The central box corresponds to ${\bf n}=0$. 
Dynamical variables such atomic coordinates, momenta, and per-atom energies  are specified by the atomic number $i$ and the vector ${\bf n}$.
In particular, atomic coordinates are given by
\be
 {\bf r}_{i{\bf n}}={\bf r}_{i}+{\bf n},  \label{coordinate}
 \ee
 where ${\bf r}_{i}$ is the coordinate of atom $i$ in the central box.
When deriving the expression for the heat current and using Hamilton's equations, the coordinates and momenta of atoms in different boxes are treated as independent variables 
with the proper periodicity for these variables imposed in the final expressions.
In our notation we drop 
subscript ${\bf n}=0$ when referring to coordinates, momenta, velocities, and other atomic properties of the central box, i. e. {${\bf r}_{i0}={\bf r}_{i}$, etc.

The heat current expression can be derived from the equation for energy conservation in the local form \cite{Hardy}
\be
\frac{\partial e(\bf{r}) }{\partial t}=-\nabla \cdot \bf{j}(\bf{r}). \label{one}
\ee
Here $e(\bf{r})$ and $\bf{j}(\bf{r})$ are, respectively, the energy density and energy flux at point $\bf{r}$. 
The microscopic energy density of a system of $N$ atoms replicated in all directions to infinity is given by 
\be 
e({\bf{r}})=\sum_{\bf n}\sum_{i=1}^N\varepsilon_{i{\bf n}}\delta({\bf{r}}-{\bf r}_{i{\bf n}}), \label{conserv}
\ee
where $\varepsilon_{i{\bf n}}$ is the atomic energy of atom $i$ in box ${\bf n}$  given
by the sum of the atomic kinetic and potential energies,
\be
\varepsilon_{i{\bf n}}=\varepsilon_{i{\bf n}}^{\text {kin}}+u_{i{\bf n}}, \label{parten}
\ee
where
\be
\varepsilon_{i{\bf n}}^{\text {kin}}=\frac{|{\bf p}_{i{\bf n}}|^2}{2m_{i{\bf n}}}, \label{kinen}
\ee
and  ${\bf p}_{i{\bf n}}$ and $m_{i{\bf n}}$ are, respectively, the momentum and mass of atom $i$ in box ${\bf n}$. The atomic potential energy $u_{i{\bf n}}$ in Eq. (\ref{parten}), which represents a given atom contribution to the total potential energy, is, 
in general, a function of atomic coordinates both in box ${\bf n}$ and other boxes. The summation over ${\bf n}$ in
 Eq. (\ref{conserv})
 runs over all values of vector ${\bf n}$ as given in Eq. (\ref{vector}).
 The total heat current for all the boxes ${\bf J}^{\text{boxes}}$ is given by the integral of the energy flux over volume, ${\bf J}^{\text{boxes}}=\int_Vd^3{\bf r}j({\bf r})$.
 To derive an explicit expression for ${\bf J}^{\text{boxes}}$  we multiply both sides of Eq. (\ref{one}) by $\bf{r}$ and integrate over volume {\cite{Baroni}}.
\be
\int_Vd ^3{\bf{r}} \,\,{\bf{r}} \frac{\partial e({\bf{r}}) }{\partial t}=-\int_Vd^3{\bf{r}} \,{ \bf{r}}\,\nabla \cdot \bf{j}(\bf{r}).
\ee
Using the energy density definition (\ref{conserv}) and performing integrations by parts we obtain
\bea
{\bf J}^{\text{boxes}}&=&\frac{d}{dt}\sum_{\bf n}\sum_{i=1}^N\varepsilon_{i \bf n}{\bf r}_{i \bf n} \\ &=&\sum_{\bf n}\sum_{i=1}^N\left(\varepsilon_{i \bf n}\frac{d{\bf r}_{i \bf n}}{dt}+{\bf r}_{i \bf n}
\frac{d\varepsilon_{i \bf n}}{dt}\right). \label{total}\
\eea
We now apply Hamilton's equations to eliminate the time derivatives. The first term in parentheses in Eq. (\ref{total}) corresponds to the convective heat current for all boxes. It can be 
written as
\be
\sum_{\bf n}\sum_{i=1}^N\varepsilon_{i \bf n}\frac{d{\bf r}_{i \bf n}}{dt}=\sum_{\bf n}\sum_{i=1}^N\varepsilon_{i \bf n}\frac{{\bf p}_{i \bf n}}{m_{i{\bf n}}}=\sum_{\bf n}\sum_{i=1}^N\varepsilon_{i}{\bf v}_{i}, \label{conv}
\ee
where, to get the last expression, we re-expressed momenta in terms of velocities ${\bf v}_{i}$ (that are commonly used in molecular dynamics) and used the fact that 
${\bf v}_{i{\bf n}}={\bf v}_{i}$ and ${\varepsilon}_{i{\bf n}}={\varepsilon}_{i}$. 
It follows from Eq. (\ref{conv}) that the convective current for one box, ${\bf J}^{\text {conv}}$ is
\be
{\bf J}^{\text {conv}}=\sum_{i=1}^N\varepsilon_{i}{\bf v}_{i}. \label{convone}
\ee

To evaluate the second term in parentheses in Eq. (\ref{total}) we use time derivatives of the local kinetic and potential energies
\bea
\frac{d\varepsilon_{i \bf n}^{\text{kin}}}{dt} &=&
-\sum_{{\bf m}}\sum_{j=1}^N\frac{\partial u_{j{\bf m}}}{\partial {\bf r}_{i \bf n}} \cdot \frac{{\bf p}_{i \bf n}}{m_{i \bf n}}, \label{ktd} \\
\frac{du_{i \bf n}}{dt} &=&
\sum_{{\bf m}}\sum_{j=1}^N\frac{\partial u_{i{\bf n }}}{\partial {\bf r}_{j \bf m}} \cdot \frac{{\bf p}_{j \bf m}}{m_{j \bf m}}. \label{ptd}
\eea
In these expressions $(-\partial u_{j\bf m}/\partial {\bf r}_{i\bf n})$ represents the partial force on atom $i$ in the ${\bf n}$th box due to the potential energy
of atom $j$ located in the ${\bf m}$th box.
Using Eqs. (\ref{ktd}) and (\ref{ptd}) and applying some summation variable changes we obtain 
\bea
& &\sum_{\bf n}\sum_{i=1}^N{\bf r}_{i \bf n}\frac{d\varepsilon_{i \bf n}}{dt} =
-\sum_{{\bf n},{\bf m}}\sum_{i,j=1}^N
({\bf r}_{i \bf n}-{\bf r}_{j \bf m})\left(\frac{\partial u_{j{\bf m}}}{\partial {\bf r}_{i \bf n}} \cdot {\bf v}_i\right) \non \\ \label{deriva}
& & =
-\sum_{{\bf n},{\bf m}}\sum_{i,j=1}^N
\big({\bf r}_{i}-{\bf r}_{j}-({\bf m}-{\bf n})\big)\left(\frac{\partial u_{j({\bf m}-{\bf n})}}{\partial {\bf r}_{i}} \cdot {\bf v}_i\right) \non \\
& & =-\sum_{{\bf n},{\bf m}}\sum_{i,j=1}^N
\left({\bf r}_{i}-{\bf r}_{j}-{\bf m}\right)\left(\frac{\partial u_{j{\bf m}}}{\partial {\bf r}_{i}} \cdot {\bf v}_i\right). \label{third}
\eea
Here, to obtain the second line, we used Eq. (\ref{coordinate}) and the fact that  because of the periodicity of the whole system $\partial u_{j{\bf m}}/\partial {\bf r}_{i \bf n}$ depends on ${\bf m}$ and ${\bf n}$ 
only through their difference ${\bf m}-{\bf n}$ and, to obtain the last 
equality, we used ${\bf m}-{\bf n}$ as the new summation variable.
The last expression represents the non-convective current for all boxes. It involves the summation over ${\bf n}$  with the summand that does not depend 
on ${\bf n}$. Therefore, the non-convective current for one (central) box, ${\bf{J}}^{\text {nconv}}$,  is 
\be
{\bf{J}}^{\text{nconv}}=\sum_{{\bf m}}\sum_{i,j=1}^N\left({\bf r}_i-{\bf r}_{j{\bf m}}\right)\left(\frac{\partial u_{j\bf m}}{\partial {\bf r}_{i}}\cdot{\bf v}_i\right). \label{nonconv}
\ee
Combining Eqs. (\ref{convone}) and (\ref{nonconv}) we obtain the sought expression for the total heat current of the system of $N$ atoms subject to periodic boundary conditions,
\be
{\bf J}=\sum_{i=1}^N\varepsilon_i{\bf v}_i-\sum_{{\bf m}}\sum_{i,j=1}^N\left({\bf r}_i-{\bf r}_{j{\bf m}}\right)\left(\frac{\partial u_{j\bf m}}{\partial {\bf r}_{i}}\cdot{\bf v}_i\right). \label{totalone}
\ee

Let us compare the last expression to the heat current for a nonperiodic system given by 
\be
{\bf J}^{\text {bulk}}=\frac{d}{dt}\sum_{i=1}^N\varepsilon_i{\bf r}_i, \label{bul}
\ee
which we will refer to as the bulk current. (Note that $\varepsilon_i$ here still depend on atomic coordinates for the image boxes but ${\bf r}_i$ is restricted to the central box.)
 One can verify
using Hamilton's equations that 
\be
{\bf J}^{\text {bulk}}=\sum_{i=1}^N\varepsilon_i{\bf v}_i-\sum_{{\bf m}}\sum_{i,j=1}^N({\bf r}_i-{\bf r}_{j})\left(\frac{\partial u_{j\bf m}}{\partial {\bf r}_{i}}\cdot{\bf v}_i\right). \label{bulk}
\ee
Comparing this to Eq. (\ref{totalone}) one can see that the total heat current for a periodic system can be written as
\be
{\bf J}={\bf J}^{\text {bulk}}+{\bf J}^{\text{bound}},  \label{percur}
\ee
where ${\bf J}^{\text{bound}}$, which we will refer to as the boundary current, is given by
\be
{\bf J}^{\text{bound}}=\sum_{{\bf m}}\sum_{i,j=1}^N{\bf m}\left(\frac{\partial u_{j \bf m}}{\partial {\bf r}_{i}}\cdot{\bf v}_i\right). \label{jbou}
\ee
It represents a contribution to the total heat current due to the energy transfer through the periodic boundaries.
This expression can be viewed in two different ways. First, up to a sign, $\sum_{j=1}^N\partial u_{j \bf m}/\partial {\bf r}_{i}$ in Eq. (\ref{jbou}) represents the sum of partial forces on atom $i$ in the central box from all the atoms in box ${\bf m}$.
 Second, using Hamilton's equations Eq. (\ref{jbou}) can also be written in the following form
\be
{\bf J}^{\text{bound}}=-\frac{1}{2}\sum_{{\bf m}}\sum_{i,j=1}^N{\bf m}\left\{\varepsilon_i,\varepsilon_{j{\bf m}}\right\}=-\frac{1}{2}\sum_{{\bf m}}{\bf m}\left\{E,E_{\bf m}\right\},
\ee
where $E_{\bf m}=\sum_{j=1}^N\varepsilon_{j{\bf m}}$ and $\{X,Y\}$ denotes the Poisson bracket of dynamical variables $X$ and $Y$. Each of the Poisson brackets 
$\left\{E,E_{\bf m}\right\}$ can be interpreted as the change in the central box energy due to its interaction with box ${\bf m}$. This interpretation is compatible with the 
energy conservation for the central box: Because of the system periodicity $\left\{E,E_{\bf m}\right\}=-\left\{E,E_{-\bf m}\right\}$, i. e. changes in the central box energy due to its interaction with boxes ${\bf m}$ and $-{\bf m}$ have the same magnitude but opposite signs and the energy flow from the central box to box ${\bf m}$
is the same as the energy flow from  box $-{\bf m}$ to the cental box.

Writing the total current as a sum of two terms in Eq. (\ref{percur}) is very similar to the separation of the virial tensor for the periodic systems \cite{Bekker,Thompson}, which 
can be written as a sum of the nonperiodic part and the boundary contribution. To make this connection more explicit
we rewrite Eq. (\ref{totalone}) as
\be
{\bf J}=\sum_{i=1}^N\varepsilon_i{\bf v}_i-\sum_{i=1}^N{\bf S}_{i}\cdot{\bf v}_i, \label{totalones}
\ee
where we introduced the atomic stress tensor
\be
{\bf S}_{i}=\sum_{{\bf m}}\sum_{j=1}^N\left({\bf r}_i-{\bf r}_{j{\bf m}}\right)\otimes\frac{\partial u_{j\bf m}}{\partial {\bf r}_{i}}. \label{totals}
\ee
The atomic stress tensor can be written as
\be
{\bf S}_{i}={\bf S}_{i}^{\text{bulk}}+{\bf S}_{i}^{\text{bound}}, \label{ts}
\ee
where
\be
{\bf S}_{i}^{\text{bulk}}=\sum_{{\bf m}}\sum_{j=1}^N({\bf r}_i-{\bf r}_{j})\otimes\frac{\partial u_{j\bf m}}{\partial {\bf r}_{i}}
\ee
and
\be
{\bf S}_{i}^{\text{bound}}=-\sum_{{\bf m}}\sum_{j=1}^N{\bf m}\otimes\frac{\partial u_{j\bf m}}{\partial {\bf r}_{i}}.
\ee
Thus, the partitioning of the total current ${\bf J}$ as in Eq. (\ref{percur}) can be reduced to the partioning of the atomic stress tensor given by Eq. (\ref{ts}),
which is the per-atom generalization of the total virial tensor separation into the bulk (nonperiodic) and boundary terms as discussed in Refs. \cite{Bekker,Thompson}.

For practical applications of Eqs. (\ref{totalone},\ref{totalones}) one needs a more specific definition of the atomic potential energy $u_{j\bf n}$. 
The system potential energy is usually given by a sum of two-atom, three-atom, and, in general, higher-order few-atom interaction terms. 
Each of these few-atom terms is commonly referred to as a group \cite{Thompson,Surblys}.
To define the atomic potential energy the 
potential energy of each group is divided equally among the atoms in that group.
The atomic potential energy is then composed of these fractional energies from all the groups which involve that atom.
This definition is not unique but there is both numerical  \cite{Schelling,Ercole} and theoretical \cite{Ercole,Ercole2016} evidence that 
the exact way by which the system potential energy is split 
into atomic potential energies does not affect the value of the thermal conductivity tensor.
 With the definition of the atomic potential energy given above, the summations over $j$ and ${\bf m}$ in Eq. (\ref{totals}) can be performed after which one is left with the summation over all the groups that involve atom $i$. After some term rearrangement and relabelling we 
 can rewrite the atomic stress in Eq. (\ref{totals}) as
 \be
 {\bf S}_i=-\sum_{k_i=1}^{K_i}({\bf r}_i-{\overline {\bf r}_{k_i}})\otimes {\bf f}_{i}^{k_i}. \label{surb}
 \ee
 Here integer index $k_i$ labels one of the groups that involve atom $i$, $K_i$ is the total number such groups and the summation in Eq. (\ref{surb})
 runs over all these groups; ${\bf f}_{i}^{k_i} $ is the partial force on atom $i$ from group $k_i$ and ${\overline {\bf r}_{k_i}}$ is the geometric center (or centroid) of the group $k_i$ given by the arithmetic average of coordinates of all atoms in the group. The last expression for the atomic stress tensor is equivalent to the one
 obtained by \textcite{Surblys} although it has a slightly different form: ${\bf r}_i$ in Eq. (\ref{surb}) is always restricted to the central box (so that all atoms are treated on an equal basis) whereas in the expression
 given in \cite{Surblys} ${\bf r}_i$ can be either in the local box or one of the image boxes depending on the group.
 
In the case of pairwise potentials an alternative but equivalent form of Eq. (\ref{surb}) that follows directly from (\ref{totals}) and can be easier to apply is
\be
 {\bf S}_i=-\frac{1}{2}\sum_{\bf m}\mathop{\sum_{j=1}^N}_{j\neq i}({\bf r}_i- {\bf r}_{j{\bf m}})\otimes {\bf F}_{ij{\bf m}}. \label{ppp}
\ee
Here ${\bf F}_{ij{\bf m}}$  the force on atom $i$ from atom $j$ located in the ${\bf m}$th box. 
The bulk and boundary contributions to the atomic stress tensor (\ref{ppp}) have the form
\bea
& &{\bf S}_i^{\text{bulk}}=-\frac{1}{2}\sum_{\bf m}\mathop{\sum_{j=1}^N}_{j\neq i}({\bf r}_i- {\bf r}_{j})\otimes {\bf F}_{ij{\bf m}}, \label{bu} \\
& &{\bf S}_i^{\text{bound}}=\frac{1}{2}\sum_{\bf m}\mathop{\sum_{j=1}^N}_{j\neq i}{\bf m}\otimes {\bf F}_{ij{\bf m}}. \label{bo}
\eea
These can be used to partition the total heat current into the bulk and boundary components for systems with pairwise potentials.

\section{Implications of the heat current definition for simulations of thermal conductivity in solids} \label{Section3}.

In the previous section we showed that the heat current can be written as a sum of two terms as in Eq. (\ref{percur}).
The bulk term of the heat current given by Eq. (\ref{bul}) has the form of the time derivative of function $ \sum_{i=1}^N\varepsilon_i{\bf r}_i $.
For finite-size solids (such as the ones considered in molecular dynamics simulations) this function is a bounded function of time. Indeed, for a given $i$, $\varepsilon_i$ cannot exceed the total system energy which is finite and  ${\bf r}_i$  oscillates about its equilibrium position specified by a finite size vector. It was shown in Refs. \cite{Baroni,Baroni2,Ercole,Ercole2016}
that adding terms that are time derivatives of bounded functions of time to the heat current does not affect the thermal conductivity calculated using the Green-Kubo approach. For completeness we demonstrate this in the Appendix where some additional conditions on these bounded functions are also given. 
Therefore the boundary term ${\bf J}^{\text{bound}}$ alone can be expected to give the same thermal conductivity as the total heat current (\ref{totalone}) when used in the Green-Kubo expression.
We verify this fact numerically  in the next section.

The use of  ${\bf J}^{\text{bound}}$ instead of ${\bf J}$ has another implication. Translating the periodic box
along {\bf a}, {\bf b}, or {\bf c} does not affect atomic dynamics and does not change the total heat current ${\bf J}$.
 However, such translation changes both ${\bf J}^{\text{bulk}}$
and ${\bf J}^{\text{bound}}$ as both of these depend on the location of boundary, i. e. the same  ${\bf J}$ can be partitioned into different ${\bf J}^{\text{bulk}}$ and ${\bf J}^{\text{bound}}$ that depend on where the boundary is placed. We can now apply the reasoning used just above to the new ${\bf J}^{\text{bulk}}$
and ${\bf J}^{\text{bound}}$ and conclude that ${\bf J}^{\text{bound}}$ associated with {{\it any}} boundary of the simulation box obtained by translating the original box along {\bf a}, {\bf b}, or {\bf c}  by an arbitrary distance can be used to calculate thermal conductivity in solids. 

An important comment has to be made regarding the validity of the arguments in the preceding two paragraphs if boundary crossings occur during system dynamics.
By examining Eqs. (\ref{totalone},\ref{bulk},\ref{jbou}) one can verify that whereas the expression for the full current (\ref{totalone}) remains a differentiable function of time, both ${\bf J}^{\text{bulk}}$ and  ${\bf J}^{\text{bound}}$ undergo step-function-like jumps when atoms cross periodic boundaries.
In solids, with atoms moving in the vicinity of the their equilibrium positions, there may or may not be boundary crossings depending on the system and the choice of the boundary. If the boundary crossings 
occur ${\bf J}^{\text{bulk}}$ does not represent a time derivative of
a bounded function.
However, the boundary crossing discontinuities for ${\bf J}^{\text{bulk}}$ and  ${\bf J}^{\text{bound}}$ can be avoided by modifying the way atoms are assigned to boxes: Each atom can be assigned to the same box it occupies at $t=0$ even if it crosses the boundary later, with the unwrapped coordinates used to specify that atom position. This approach ensures continuity of ${\bf J}^{\text{bulk}}$ and  ${\bf J}^{\text{bound}}$ as functions of time without changing the value of the full current (\ref{totalone}). An alternative view of this procedure is to assume that as the periodic system evolves in time the original 
parallelepiped boxes are  deforming (possibly in a complex way) so that atoms do not cross the boundaries of these boxes. With atoms never crossing the boundaries, atomic coordinates remain continuous functions of time and ${\bf J}^{\text{bulk}}$ is a time derivative of a bounded function and, therefore, ${\bf J}^{\text{bound}}$ should lead to the same thermal conductivity as the total ${\bf J}$.

\section{Numerical demonstration} \label{two}
In this section we verify the conclusions that we reached  in the preceding  section.  We consider two systems: a single crystal of argon and  crystals of argon and krypton forming an interface. Both systems are studied using molecular dynamics. 
Both argon and krypton are modeled using the pairwise Lennard-Jones potential 
\be
V(\rho)=4\epsilon\left[\left(\frac{\sigma}{\rho}\right)^{12}-\left(\frac{\sigma}{\rho}\right)^{6}\right] \label{LJ},
\ee
where $\rho$ is the distance between the two atoms and $\sigma$ and $\epsilon$ are parameters specifying the potential.
The parameters for argon were chosen to be $\sigma_{\text {Ar}}=3.401$ {\AA} and $\epsilon_{\text {Ar}}=0.2339$ kcal/mol \cite{OPLS} and for krypton $\sigma_{\text {Kr}}=3.591$ {\AA} and $\epsilon_{\text {Kr}}=0.34319$ kcal/mol  \cite{Beattie}.
In the case of the argon-krypton interaction the Lorentz-Berthelot combining rules were used, i. e. $\sigma_{\text {ArKr}}=(\sigma_{\text {Ar}}+\sigma_{\text {Kr}})/2$ and  $\epsilon_{\text {ArKr}}=\sqrt{\epsilon_{\text {Ar}}\epsilon_{\text {Kr}}}$. A cutoff distance of 11 {\AA} was used both for argon and krypton. The time step was set to 2 fs. Orthogonal simulation boxes with 3-D periodic boundary conditions were used. All simulations were performed using LAMMPS \cite{Plimpton}.

In the first set of simulations a single argon crystal consisting of $5\times5\times5$ conventional face-centered cubic unit cells (500 Ar atoms). Figure \ref{Figure0} shows the general setup of the simulation cell.
\begin{figure}
 \includegraphics[width=\columnwidth]{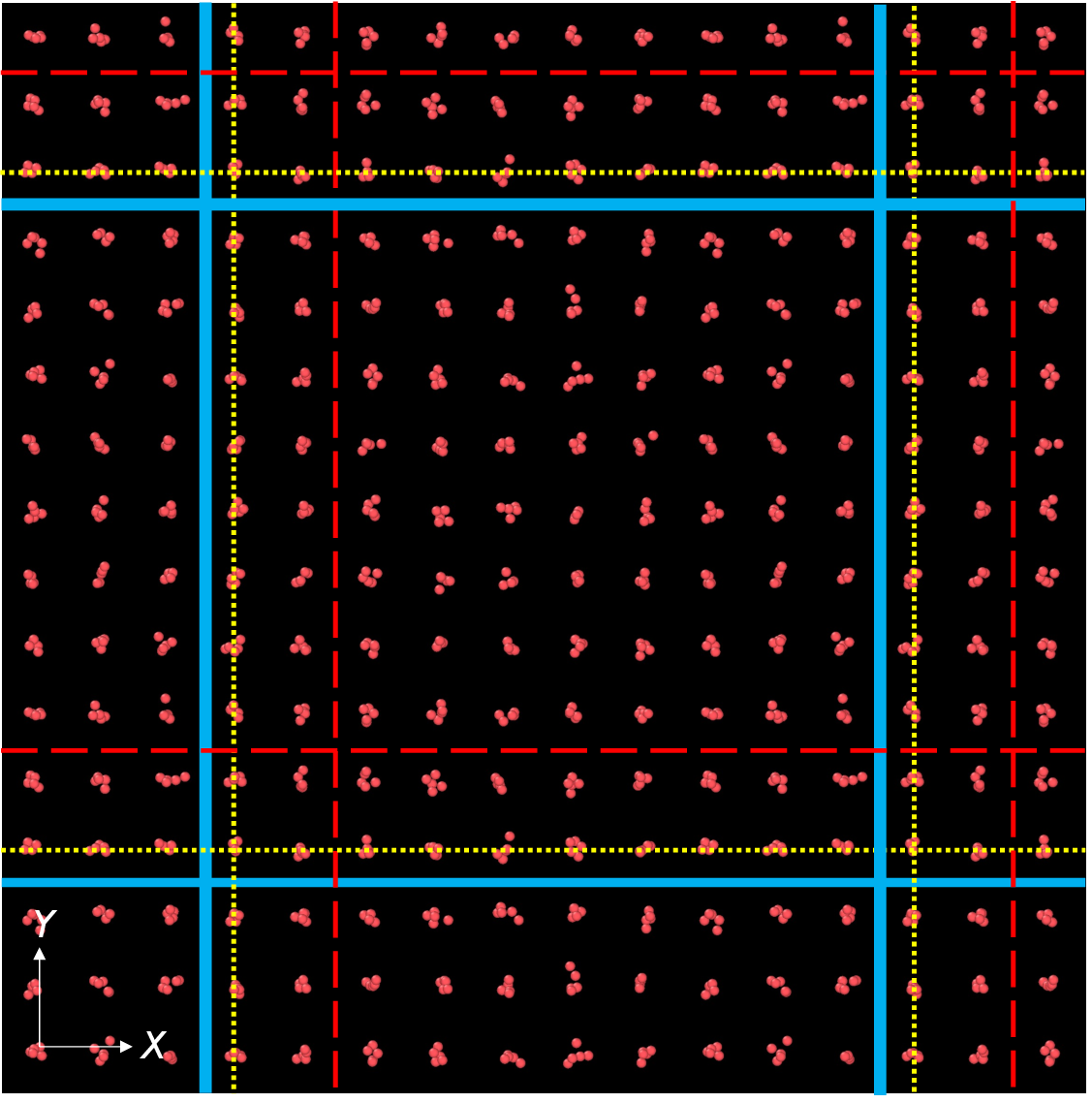}  
 \caption{\label{Figure0} Projection of the  Ar crystal surrounded by its periodic images onto the $xy$-plane. 
  The three sets of boundaries considered are shown with solid blues lines, red dashes, and yellow dots.}
 \end{figure}
The system was equilibrated at temperature $T=50$ K and pressure $P=1$ atm using an NPT ensemble to obtain the average lattice parameters. The resulting cubic simulation box was
26.89893 {\AA}$\times$26.89893 {\AA}$\times$26.89893 {\AA} along $x$, $y$, and $z$ axes. A production equilibrium NVE trajectory was run for 80 ns. Atomic coordinates, velocities, and heat current components  were recorded every 20 fs. Forces between individual  atomic pairs   were calculated using the derivatives of the Lennard-Jones potential (\ref{LJ}) and atomic coordinates. Using these data, the bulk and boundary contributions to the total heat current were calculated (cf. Eqs. (\ref{bu},\ref{bo})) as functions of time for the boundary shown in blue in Fig. \ref{Figure0}.
For this  choice of boundary conditions boundary crossings never occurred.  
The calculated  ${\bf J}^{\text {bulk}}$, ${\bf J}^{\text {bound}}$, and the total heat current ${\bf J}$ were used to obtain the corresponding current correlation functions and thermal conductivities. The results of this analysis are summarized in Figs. \ref{Figure1} and \ref{Figure2}. 
 \begin{figure}
 \includegraphics[width=\columnwidth]{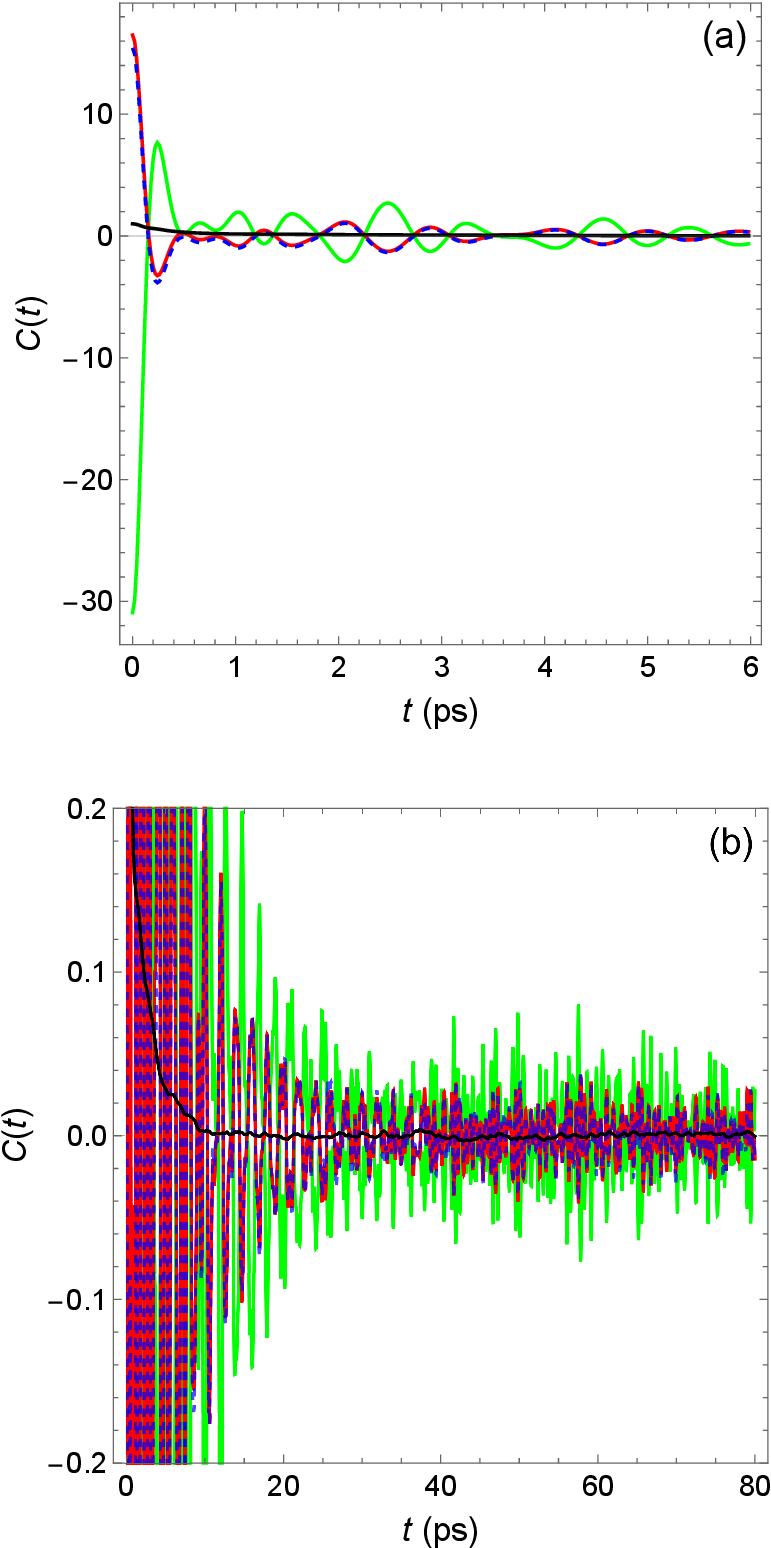}  
 \caption{\label{Figure1}  (a) Current correlation functions $C^{\text {tot}}(t)$  (black), 
 $C^{\text {bulk}}(t)$ (blue dashes), 
 $C^{\text {bound}}(t)$ (red), and $C^{\text{cross}}(t)$ (green). 
 (b) Same as (a) but for 
 longer times with a finer scale for the $y$-axis.}
 \end{figure}
Figure \ref{Figure1} shows the current correlation functions for total heat current $C^{\text {tot}}(t)$,  bulk current $C^{\text {bulk}}(t)$,  boundary current $C^{\text {bound}}(t)$, and the bulk-boundary cross-correlation function $C^{\text {cross}}(t)$. In each case, the average of the almost identical $xx$, $yy$, and $zz$ correlation function components is shown.
All correlation functions were scaled and made dimensionless by ensuring
 that $C^{\text {tot}}(0)$ is equal to unity.  Note that  $C^{\text {tot}}(t)=C^{\text {bulk}}(t)+C^{\text {bound}}(t)+C^{\text {cross}}(t)$. One can see that at $t=0$ the magnitude of   $C^{\text {tot}}(t)$ is much smaller than the magnitudes for the other three correlation functions. All four functions in Fig. \ref{Figure1} decay to zero but in different ways. Whereas $C^{\text {tot}}(t)$ is a monotonically decreasing function of time until it essentially decays,  $C^{\text {bulk}}(t)$, $C^{\text {bound}}(t)$, and $C^{\text {cross}}(t)$  exhibit oscillatory decay. $C^{\text {bulk}}(t)$ and  $C^{\text {bound}}(t)$ behave very similarly for the entire time interval considered.  
 All four functions fluctuate about zero for longer times but the magnitude of these fluctuations for $C^{\text {tot}}(t)$ is much smaller than those for  $C^{\text {bulk}}(t)$, $C^{\text {bound}}(t)$, and
  $C^{\text {cross}}(t)$. The magnitude of the long-time fluctuations for $C^{\text {cross}}(t)$ is about twice that for $C^{\text {bulk}}(t)$ and $C^{\text {bound}}(t)$.
The time-dependent thermal conductivity for each $C(t)$  is defined as
$\kappa(t)=K\int_0^t d\tau C(\tau)$, where $K$ is a $T$ and $V$ dependent parameter that ensures correct units for $\kappa$.
The thermal conductivities $\kappa^{\text {tot}}(t)$, $\kappa^{\text {bulk}}(t)$, $\kappa^{\text {bound}}(t)$, and $\kappa^{\text {cross}}(t)$ obtained, respectively, 
from $C^{\text {tot}}(t)$, $C^{\text {bulk}}(t)$, $C^{\text {bound}}(t)$, and $C^{\text {cross}}(t)$ 
are shown in Fig. \ref{Figure2}.
 \begin{figure}
 \includegraphics[width=\columnwidth]{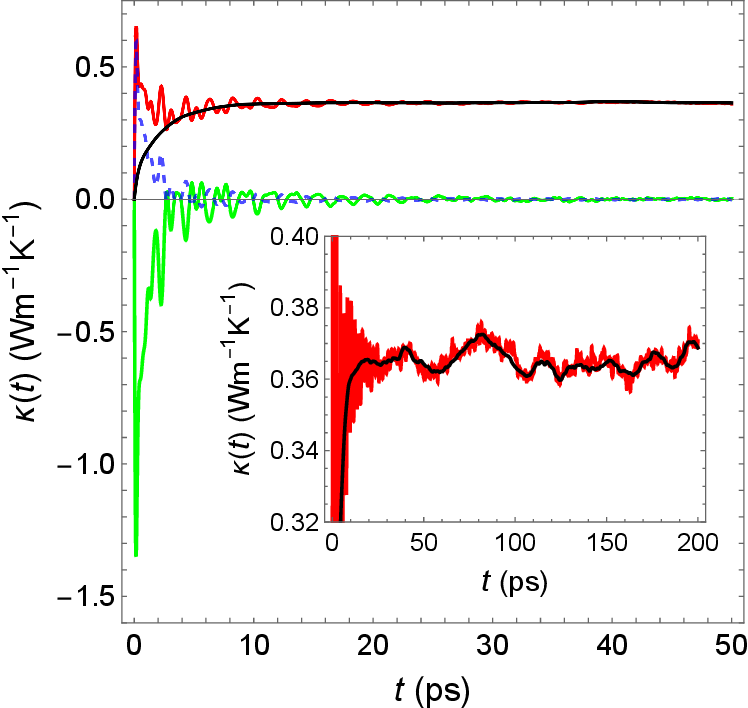}  
 \caption{\label{Figure2}  Time-dependent thermal conductivities $\kappa^{\text {tot}}(t)$ (black), $\kappa^{\text {bulk}}(t)$ (blue dashes), $\kappa^{\text {bound}}(t)$ (red), and $\kappa^{\text {cross}}(t)$ (green) obtained from the four correlation functions shown in Fig. \ref{Figure1}.
 The inset shows $\kappa^{\text {tot}}(t)$ (black) and 
 $\kappa^{\text {bound}}(t)$ (red) for longer times with a finer scale for the $y$-axis.}
 \end{figure}
One can see that, as anticipated, $\kappa^{\text {bulk}}(t)$ and $\kappa^{\text {cross}}(t)$ decay to zero as functions of time whereas both $\kappa^{\text {tot}}(t)$ and $\kappa^{\text {bound}}(t)$ converge to the same approximately constant positive value. The inset of Fig. \ref{Figure2} shows that 
 $\kappa^{\text {tot}}(t)$ and $\kappa^{\text {bound}}(t)$ behave similarly for longer times as well, although $\kappa^{\text {bound}}(t)$ exhibits more fluctuations.
 Thus, both ${\bf J}$ and ${\bf J}^{\text {bound}}$ do give the same value for the thermal conductivity when used in the Green-Kubo expression.
 
Next we verified that ${\bf J}^{\text {bound}}$ calculated for other boundaries obtained by translating the original boundary gives the same thermal conductivity. To this end we used the same NVE trajectory and calculated ${\bf J}^{\text {bound}}(t)$ for boundaries that are formed by shifting the original boundary shown in blue in Fig. \ref{Figure0} by distance $a$ simultaneously along $x$, $y$, and $z$ axes. We considered two values of $a$: 5.379786 {\AA} and 1.3 {\AA} (with the corresponding  boundaries shown with red dashes and yellow dots in Fig. \ref{Figure0}). The first choice of $a$ corresponds to the length of 
  the unit cell lattice vector and the second one ensures that there are boundary crossings as the system evolves in time. The boundary crossings for the second choice of $a$ were 
  effectively eliminated by using unwrapped coordinates as discussed in Sec. \ref{Section3}. All three boundary choices correspond to the same total current ${\bf J}$.
 \begin{figure}[hbt!]
 \includegraphics[width=\columnwidth]{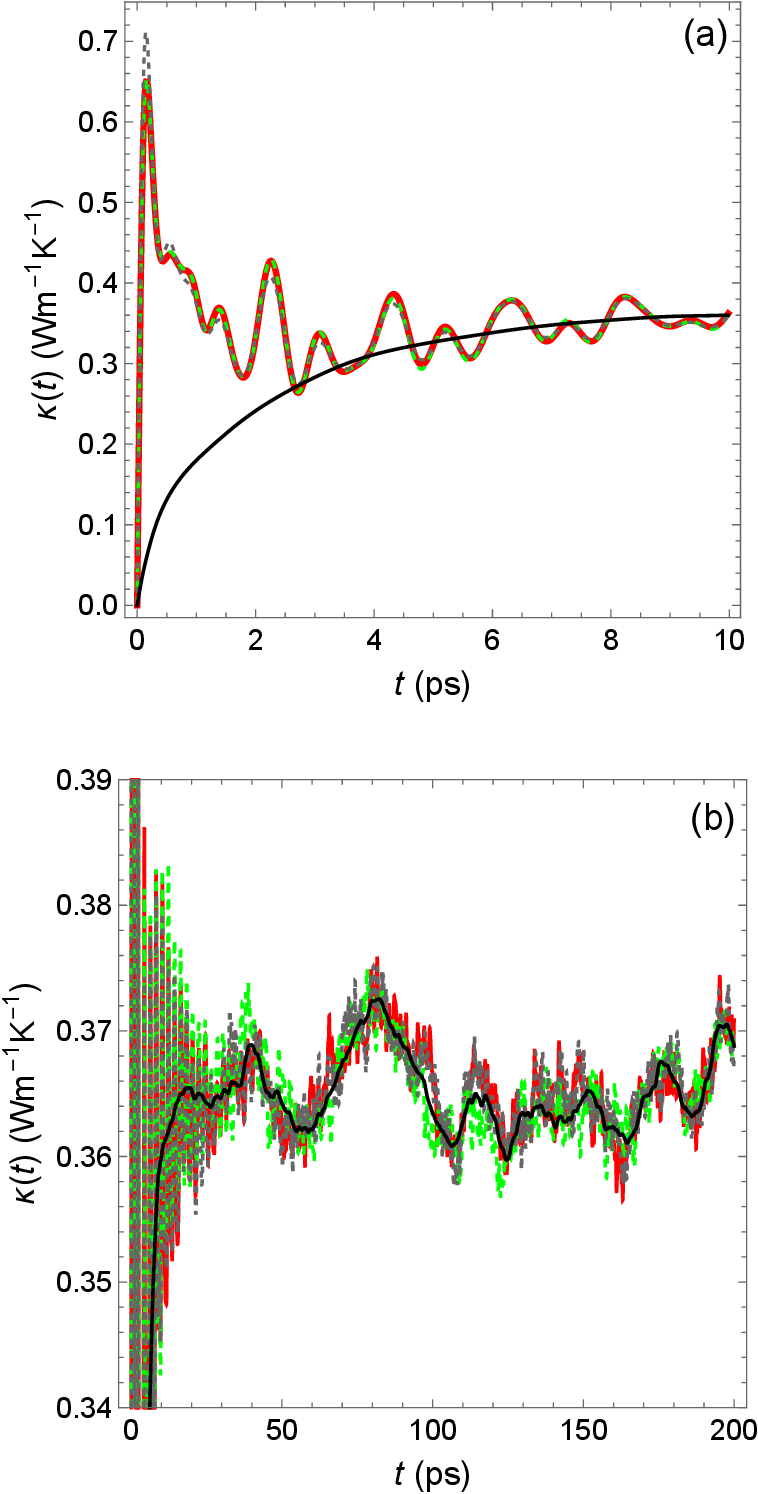}  
 \caption{\label{Figure2a} 
(a) $\kappa^{\text {bound}}(t)$'s calculated for the original boundary (red),
 the original boundary shifted by 5.379786 {\AA} (green dashes) and 1.3 {\AA} (gray dashes). $\kappa^{\text {tot}}(t)$ is shown in black. (b) 
 Same as (a) but for 
 longer times with a finer scale for the $y$-axis.}
 \end{figure} 
 The time-dependent thermal conductivities calculated from  ${\bf J}^{\text bound}(t)$ for the three boundaries  along with the conductivity obtained from the total ${\bf J}$ are shown in Fig. \ref{Figure2a}.
 One can see that all three  $\kappa^{\text {bound}}(t)$'s behave similarly and converge to $\kappa^{\text {tot}}(t)$. Similarly to the results 
 shown in Fig. \ref{Figure2}, all three $\kappa^{\text {bound}}(t)$'s exhibit more fluctuations than $\kappa^{\text {tot}}(t)$ for long times.

 In the second set of simulations we considered crystals of argon and krypton 
 forming an interface. The goal was to verify that thermal conductivity calculated using ${\bf J}^{\text {bound}}$ does not depend on the boundary 
 position even in a heterogeneous system. The general setup of the simulation cell is shown in Fig. \ref{Figure3}. 
  \begin{figure}[hbt!]
 \includegraphics[width=\columnwidth]{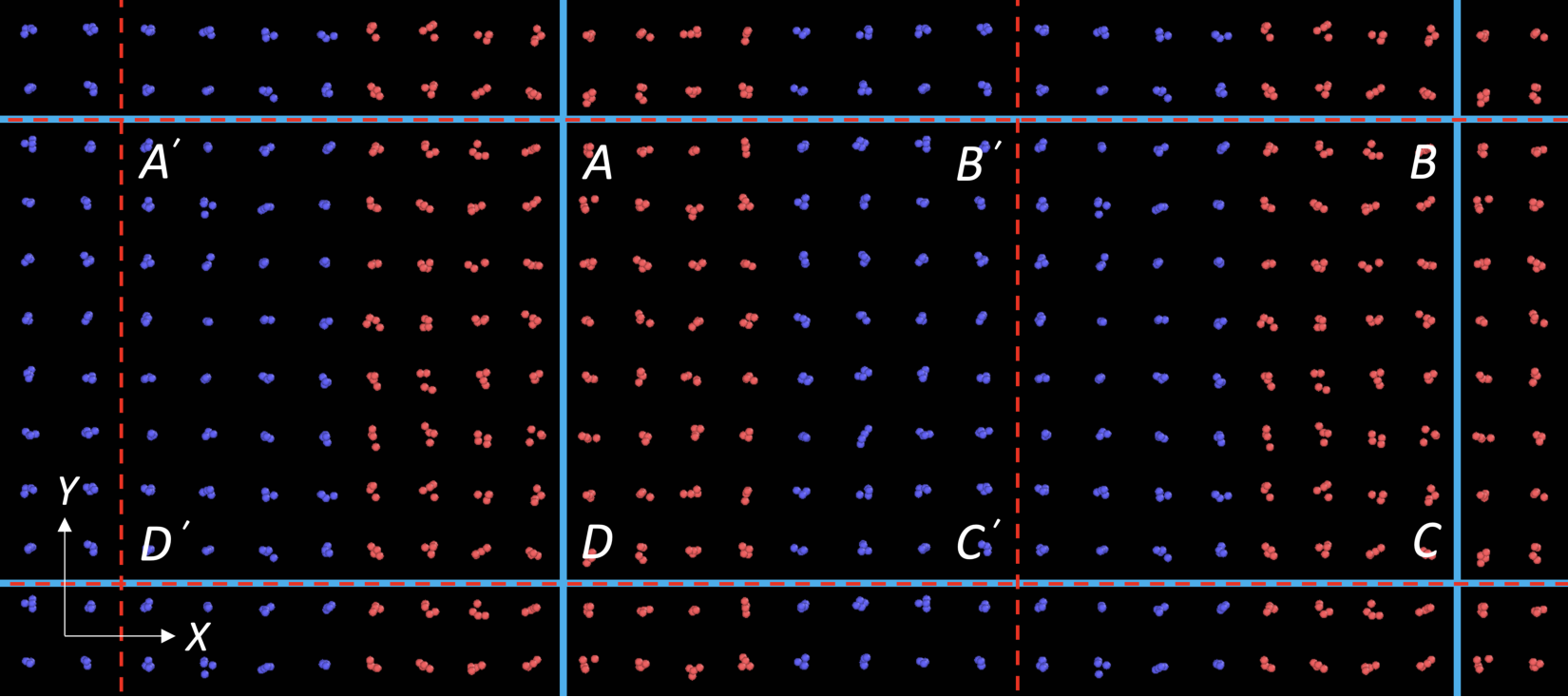}  
 \caption{\label{Figure3} Projection of the composite Ar-Kr crystal surrounded by its periodic images onto the $xy$-plane. 
 The argon atoms are red and the krypton atoms are blue. The two sets of boundaries for which $J_x^{\text {bound}}(t)$ was calculated are shown with blues lines (with the central box $ABCD$) and
 red dashes (with the central box $A'B'C'D'$).}
 \end{figure}
 Each of the two crystals consisted of  
$4\times4\times4$ conventional face-centered cubic unit cells (256 atoms). The entire system was equilibrated at 50 K and 1 atm using an NPT
 ensemble and an orthogonal simulation box to obtain the average lattice parameters. The resulting simulation box was 43.3824 {\AA}$\times22.2945$ {\AA}$\times22.2945$ {\AA} along $x$, $y$, and $z$ axes, respectively.
Then, similarly to the previous set of simulations, an 80 ns NVE trajectory corresponding to the average $T=50$ K and $P=1$ atm was run and data recorded every 20 fs.
 \begin{figure}[hbt!]
 \includegraphics[width=\columnwidth]{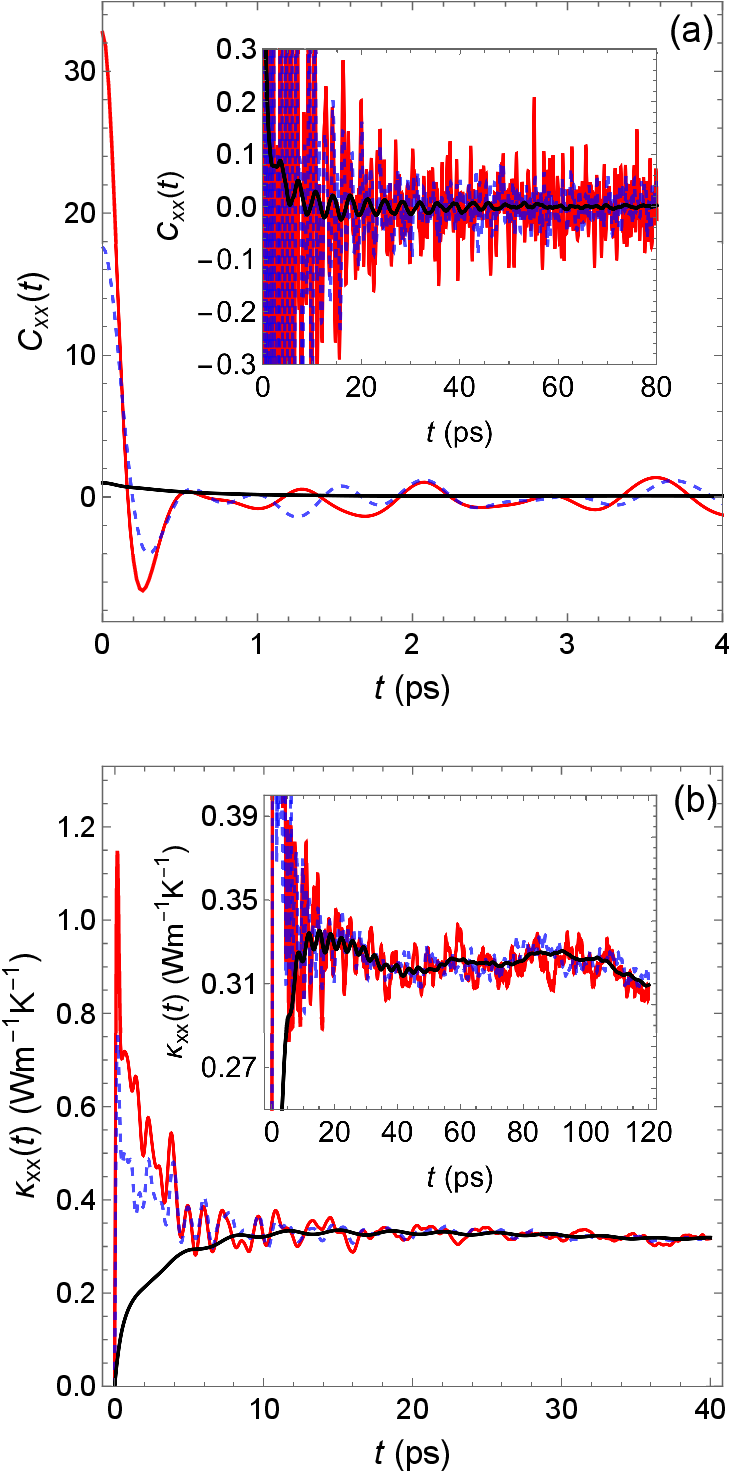}  
 \caption{\label{Figure4}  (a) The correlation functions $C_{xx}^{\text {tot}}(t)$ (black), $C_{xx}^{\text {bound,Ar}}(t)$ (red), 
 and $C_{xx}^{\text {bound,Kr}}(t)$ (blue dashes). All correlation functions are dimensionless. The inset shows the same functions for longer times with a finer scale for the $y$-axis.
 (b) The corresponding time-dependent  thermal conductivities  $\kappa_{xx}^{\text {tot}}(t)$ (black), $\kappa_{xx}^{\text {bound,Ar}}(t)$ (red), 
 and $\kappa_{xx}^{\text {bound,Kr}}(t)$ (blue dashes). The inset shows the same functions for longer times with a finer scale for the $y$-axis. }
 \end{figure}
  Here we were interested in the $xx$ component of the thermal conductivity tensor because the system is inhomogeneous  along the $x$-axis.  Two different sets of boundaries for which we chose to calculate $J_x^{\text {bound}}$
 are shown in Fig. \ref{Figure3}. The $yz$-faces of the first boundary were fully within the argon subsystem whereas they were fully within the krypton subsystem for the second boundary. The corresponding correlation functions calculated from the boundary currents are denoted $C_{xx}^{\text {bound,Ar}}(t)$ and $C_{xx}^{\text {bound,Kr}}(t)$.
 They are shown in 
Fig. \ref{Figure4}(a)  along with $C_{xx}^{\text {tot}}(t)$ obtained from the total $J_x(t)$.
Similarly to the case of the Ar crystal considered above the magnitude and general behavior of $C_{xx}^{\text {tot}}(t)$  is quite different from those for 
$C_{xx}^{\text {bound,Ar}}(t)$ and $C_{xx}^{\text {bound,Kr}}(t)$. Note that $C_{xx}^{\text {bound,Ar}}(t)$ and $C_{xx}^{\text {bound,Kr}}(t)$ also differ substantially. In particular, $C_{xx}^{\text {bound,Ar}}(t)$ is almost twice as large as $C_{xx}^{\text {bound,Kr}}(t)$ at $t=0$. $C_{xx}^{\text {bound,Ar}}(t)$ and $C_{xx}^{\text {bound,Kr}}(t)$ also differ in the way they decay.  $C_{xx}^{\text {bound,Ar}}(t)$  exhibits stronger
fluctuations around zero for long times in comparison to $C_{xx}^{\text {bound,Kr}}(t)$. 

The corresponding time-dependent thermal conductivities are shown in  Fig. \ref{Figure4}(b). In spite of the
substantial differences among the correlation functions,  all three thermal conductivities converge to the same approximately constant value. As can be seen in the inset of Fig. \ref{Figure4}(b), this also remains true for longer times. Note, however, that conductivities obtained from the boundary terms exhibit stronger fluctuations
for longer times. These results confirm that indeed any surface within a solid obtained by translating of the original boundary can be 
used to calculate ${\bf J}^{\text {bound}}$ with the corresponding thermal conductivities
obtained from different ${\bf J}^{\text {bound}}$'s and the total heat current ${\bf J}$
converging to the same value.

 \section{Conclusions}
 We re-derived an expression for the heat current for a classical system subject to periodic boundary conditions. 
 We showed that the current can be separated into two parts: the time derivative of a bounded function of time and the boundary term. 
 For finite solids both the total current and the boundary term give the same thermal conductivity when used within the Green-Kubo approach. 
 It is of interest to investigate further if using the boundary term instead of the full current can give any numerical advantages. 
 For the simple, small systems modeled with the Lennard-Jones potential 
considered in this study, using the full current yields less noisy values for the integrated thermal conductivities. However, it is 
not clear whether the same will be true for solids modeled with other potentials or when using larger simulation cells. Apart from the numerical aspect, the use of ${\bf J}^{\text {bound}}$ instead of ${\bf J}$  offers an alternative view of the thermal  transport in solids in that the thermal conductivity can be calculated from the energy passing 
through a surface per unit of time, a quantity that is local in one dimension. 
 
 \begin{acknowledgments}
This research was funded by Air Force Office of Scientific Research grants number FA9550-19-1-0318 and FA9550-22-1-0212.
The author would like to thank Tommy Sewell for valuable comments and suggestions.

\end{acknowledgments}
\appendix*
\section{}
In this appendix we show that, subject to certain conditions, adding a time derivative of a vector dynamical variable to the heat current will not affect the thermal conductivity tensor
calculated using Eq. (\ref{qcurr}). Here we mostly follow Ref. \cite{Ercole2016} with some generalizations. 
Consider heat current ${\bf J}$, some vector dynamical
variable ${\bf P}$, and the new heat current ${\bf J}'$ given by
 \be
{\bf J}'={\bf J}+\frac{d {\bf P}}{dt}.
\ee
The difference $D_{\alpha\beta}$ between integrals of the correlation functions 
calculated using 
${\bf J}'$ and ${\bf J}$ (cf. Eqs. (\ref{qcurr},\ref{corrfun})) is given by
\begin{widetext}
\bea
D_{\alpha\beta}&=&\int_0^\infty dt\langle J_{\alpha}'(0) J_{\beta}'(t)\rangle - \int_0^\infty dt\langle J_{\alpha}(0) J_{\beta}(t)\rangle
=\int_0^\infty dt\left\langle \left(J_{\beta}(t)+ \frac{d P_{\beta}(t)}{dt}\right)\frac{d P_{\alpha}(t)}{dt}\bigg|_{t=0} +\frac{d P_{\beta}(t)}{dt}J_{\alpha}(0)\right\rangle \non \\
&=&\int_0^\infty dt\left\langle -\frac{d P_{\alpha}(-t)}{dt}J_{\beta}(0)+\frac{d P_{\beta}(t)}{dt}J_{\alpha}(0)+ \frac{d P_{\beta}(t)}{dt}\frac{d P_{\alpha}(t)}{dt}\bigg|_{t=0}\right\rangle.  \label{A2a}
\eea
\end{widetext}
Here to obtain the first term in brackets in the second line we 
used the identity $\langle A(0)B(t)\rangle=\langle A(-t)B(0)\rangle$ which is valid 
for any dynamical variables $A$ and $B$ and equilibrium averages. Performing integrations in (\ref{A2a}) gives us
\begin{widetext}
\bea
D_{\alpha\beta}&=&\left\langle\Big(-P_{\alpha}(-\infty)+P_{\alpha}(0)\Big)J_{\beta}(0)+\Big(P_{\beta}(\infty)-P_{\beta}(0)\Big)J_{\alpha}(0)+\Big(P_{\beta}(\infty)-P_{\beta}(0)\Big)\frac{d P_{\alpha}(t)}{dt}\bigg|_{t=0}\right\rangle  \label{A3a} \\
&=&\left\langle P_{\alpha}(0)J_{\beta}(0)-P_{\beta}(0)J_{\alpha}(0)
-P_{\beta}(0)
\frac{d P_{\alpha}(t)}{dt}\bigg|_{t=0}\right\rangle \label{A3b} \\
&=&\left\langle P_{\alpha}(0)J_{\beta}(0)-P_{\beta}(0)J_{\alpha}(0)
+\frac{1}{2}\left(P_{\alpha}(0)
\frac{d P_{\beta}(t)}{dt}\bigg|_{t=0}-P_{\beta}(0)
\frac{d P_{\alpha}(t)}{dt}\bigg|_{t=0}\right)\right\rangle. \label{A3c} 
\eea
\end{widetext}
Here to go from (\ref{A3a}) to (\ref{A3b}) we used the fact that for nonintegrable systems
we can expect $\langle A(0)B(\pm\infty)\rangle=\langle A(0)\rangle\langle B(\pm\infty)\rangle$ for any dynamical variables $A$ and $B$. Recalling that in equilibrium $\left\langle {\bf J} \right\rangle=0$ and  $d\left\langle {\bf P} \right\rangle/dt=0$
one obtains (\ref{A3b}). Note that for this argument to be valid ${\bf P}(\pm \infty)$ has to remain bounded so that the decay of time correlations is not overtaken by growth of ${\bf P}(t)$. Finally, to obtain (\ref{A3c}) we used the fact that for equilibrium averages
$d\left\langle  P_{\alpha}P_{\beta} \right\rangle/dt=0$. Thus, equality of $D_{\alpha \beta}$ (as given by (\ref{A3c})) to zero is required for ${\bf J}$ 
and ${\bf J}'$ to give the 
same thermal conductivity tensor. Depending on a specific form of
${\bf P}$, this condition may or may not be satisfied. It is satisfied for ${\bf P}$ given by $-\sum_{i}\varepsilon_i{\bf r}_i$ considered in this work and for some other choices of ${\bf P}$ discussed in the literature \cite{Baroni2,PereverzevSewell}.  Note, however, that even if $D_{\alpha \beta} \neq 0$ it represents a skew-symmetric matrix.
Therefore, its contribution to the thermal conductivity tensor can be 
eliminated by symmetrising the off-diagonal components of $\kappa_{\alpha \beta}$, which is usually done when performing numerical analysis based on Eq. (\ref{qcurr}).
%

\end{document}